\RequirePackage{fix-cm}
\documentclass[pdftex,twocolumn,epjc3]{svjour3}

\usepackage{graphicx}

\usepackage[utf8]{inputenc}

\usepackage{xspace}

\usepackage{amssymb,amsmath}

\usepackage{calc}

\usepackage{microtype}

\usepackage[colorlinks,citecolor=blue,urlcolor=blue,linkcolor=blue]{hyperref}

\journalname{Eur. Phys. J. D}

\usepackage[numbers,sort&compress]{natbib}

\newcommand\tablefoot[1]{%
  \par\vspace{1ex}
  \noindent
  \begin{minipage}{\linewidth}
    \ignorespaces
    #1%
  \end{minipage}%
}
\newcommand*\tablefootmark[1]{%
  \unskip
  \hbox{\textsuperscript{\normalfont\ignorespaces#1}}%
  \,%
  \ignorespaces
}
\newcommand\tablefoottext[2]{%
  \hbox{\textsuperscript{\normalfont({\ignorespaces#1})}}%
  ~%
  \ignorespaces
  #2\ \ignorespaces%
}

\makeatletter

\newlength\@@boxwidth
\def\samebox#1#2{%
  \settowidth{\@@boxwidth}{#1}%
  \makebox[\@@boxwidth][c]{#2}}

\makeatother

\def\CO{\ensuremath{\mathrm{CO_2}}\xspace}
\def\NE{\ensuremath{\mathrm{Ne}}\xspace}
\def\SF{\ensuremath{\mathrm{SF_6}}\xspace}

\def\NeonI{867.278} 
\def\NeonII{868.980} 
\def\NeonIII{869.620} 
\def\NeonIV{869.920} 

\def\SFI{688.448} 
\def\SFII{692.082} 
\def\SFIII{694.217} 
\def\SFIV{696.296} 
\def\SFV{699.446} 
\def\SFEdge{696.998} 

\def\COI{535.334} 
\def\COII{535.582} 
\def\COIII{537.069} 
\def\COIV{537.937} 
\def\COV{538.487} 
\def\COVI{538.720} 
\def\COVII{538.908} 
\def\COVIII{539.197} 
\def\COIX{539.595} 

\renewenvironment{tabular}[1]{%
  \def\arraystretch{1.5}
  \begin{tabular*}{\hsize}{@{\extracolsep{\fill}}#1}}
  {\end{tabular*}}

\newcommand{\littleskip}{10pt}
\newcommand{\tboxr}[1]{%
  \setlength{\fboxrule}{0pt}%
  \setlength{\fboxsep}{4pt}%
  \fbox{$\left.\parbox{\littleskip}{\textnormal{{#1}}}\mkern9mu\right\}$}}

\newcommand{\tboxl}[1]{%
  \setlength{\fboxrule}{0pt}%
  \setlength{\fboxsep}{4pt}%
  \fbox{$\left\{\mkern9mu\textnormal{\parbox{\widthof{000.00}}{#1}}\mkern9mu\right.$}}
\newcommand{\tboxlrn}[1]{%
  \setlength{\fboxrule}{0pt}%
  \setlength{\fboxsep}{4pt}%
  \fbox{$\left.\mkern9mu\textnormal{\parbox{\widthof{000.00}}{#1}}\mkern9mu\right.$}}
\newcommand{\tboxrn}[1]{%
  \setlength{\fboxrule}{0pt}%
  \setlength{\fboxsep}{4pt}%
  \fbox{$\left.\parbox{\littleskip}{\textnormal{{#1}}}\mkern9mu\right.$}}
\newcommand{\tboxln}[1]{%
  \setlength{\fboxrule}{0pt}%
  \setlength{\fboxsep}{4pt}%
  \fbox{$\left.\mkern9mu\parbox{\widthof{000.00}}{\textnormal{{#1}}}\right.$}}

\begin{document}
\title{A new benchmark of soft X-ray transition energies of \NE, \CO, and \SF: paving a pathway towards ppm accuracy}

\author{J.~Stierhof\thanksref{inst:remeis,mail:stierhof}
\and S.~K\"uhn\thanksref{inst:mpik}
\and M.~Winter\thanksref{inst:fauphys,inst:neel}
\and P.~Micke\thanksref{inst:cern}
\and R.~Steinbr\"ugge\thanksref{inst:desy}
\and C.~Shah\thanksref{inst:gsfc,inst:mpik,inst:llnl}
\and N.~Hell\thanksref{inst:llnl}
\and M.~Bissinger\thanksref{inst:remeis}
\and M.~Hirsch\thanksref{inst:remeis}
\and R.~Ballhausen\thanksref{inst:remeis}
\and M.~Lang\thanksref{inst:remeis}
\and C.~Gr\"afe\thanksref{inst:remeis}
\and S.~Wipf\thanksref{inst:ujena}
\and R.~Cumbee\thanksref{inst:gsfc,inst:umaryland}
\and G.~L.~Betancourt-Martinez\thanksref{inst:irap}
\and S.~Park\thanksref{inst:unist}
\and J.~Niskanen\thanksref{inst:hzb}
\and M.~Chung\thanksref{inst:unist}
\and F.~S.~Porter\thanksref{inst:gsfc}
\and T.~St\"ohlker\thanksref{inst:ujena,inst:gsi,inst:hij}
\and T.~Pfeifer\thanksref{inst:mpik}
\and G.~V. Brown\thanksref{inst:llnl}
\and
S.~Bernitt\thanksref{inst:ujena,inst:hij,inst:gsi,inst:mpik}
\and P.~Hansmann\thanksref{inst:fauphys}
\and J.~Wilms\thanksref{inst:remeis}
\and J.~R.~{Crespo L\'opez-Urrutia}\thanksref{inst:mpik}
\and M.~A.~Leutenegger\thanksref{inst:gsfc}
}

\thankstext{mail:stierhof}{e-mail: jakob.stierhof@fau.de}

\institute{
Dr.~Karl Remeis-Observatory and Erlangen Centre for Astroparticle Physics Friedrich-Alexander-Universit\"at Erlangen-N\"urnberg, Sternwartstr.~7, 96049 Bamberg, Germany
\label{inst:remeis}
\and
Max-Planck-Institut f\"ur Kernphysik, Saupfercheckweg 1, 69117 Heidelberg, Germany
\label{inst:mpik}
\and
Institute of Theoretical Physics, Friedrich-Alexander-Universit\"at Erlangen-N\"urnberg, Staudtstr. 7/B2, 91058 Erlangen, Germany\label{inst:fauphys}
\and
Université Grenoble Alpes, CNRS, Institut NEEL, 25 rue des Martyrs BP 166, 38042 Grenoble Cedex 9, France
\label{inst:neel}
\and
CERN, 1211 Geneva 23, Switzerland
\label{inst:cern}
\and
Deutsches Elektronen-Synchrotron DESY, Notkestr.~85, 22607 Hamburg, Germany
\label{inst:desy}
\and
NASA Goddard Space Flight Center, 8800 Greenbelt Rd., Greenbelt, MD 20771, USA
\label{inst:gsfc}
\and
Lawrence Livermore National Laboratory, 7000 East Ave., Livermore, CA 94550, USA
\label{inst:llnl}
\and
Institut f\"ur Optik und Quantenelektronik, Friedrich-Schiller-Universit\"at Jena, Max-Wien-Platz 1, 07743 Jena, Germany
\label{inst:ujena}
\and
Department of Astronomy, University of Maryland, College Park, MD 20742
\label{inst:umaryland}
\and
Institut de Recherche en Astrophysique et Plan\'etologie, 9, avenue du Colonel Roche BP 44346, 31028 Toulouse Cedex 4, France
\label{inst:irap}
\and
Ulsan National Institute of Science and Technology, 50 UNIST-gil, Ulsan, South Korea
\label{inst:unist}
\and
Institute for Methods and Instrumentation in Synchrotron Radiation Research G-ISRR, Helmholtz-Zentrum Berlin f\"ur Materialien und Energie, Albert-Einstein-Strasse 15, 12489 Berlin, Germany
\label{inst:hzb}
\and
GSI Helmholtzzentrum f\"ur Schwerionenforschung, Planckstra{\ss}e 1, 64291 Darmstadt, Germany
\label{inst:gsi}
\and
Helmholtz-Institut Jena, Fr\"obelstieg 3, 07743 Jena, Germany
\label{inst:hij}
}

\date{Received 30 November 202/Accepted 23 January 2022}

\abstractdc{A key requirement for the correct interpretation of high-resolution
  X-ray spectra is that transition energies are known with high accuracy and
  precision. We investigate the K-shell features of \NE, \CO, and \SF gases, by
  measuring their photo ion-yield spectra at the BESSY II synchrotron facility
  simultaneously with the 1s--$n$p fluorescence emission of He-like ions
  produced in the Polar-X EBIT. Accurate \emph{ab initio} calculations of
  transitions in these ions provide the basis of the calibration. While the \CO
  result agrees well with previous measurements, the \SF spectrum appears
  shifted by $\sim$0.5\,eV, about twice the uncertainty of the earlier results.
  Our result for \NE shows a large departure from earlier results, but may
  suffer from larger systematic effects than our other measurements. The
  molecular spectra agree well with our results of time-dependent density
  functional theory. We find that the statistical uncertainty allows
  calibrations in the desired range of 1--10\,meV, however, systematic
  contributions still limit the uncertainty to ${\sim}$40--100\,meV, mainly due
  to the temporal stability of the monochromator energy scale. Combining our
  absolute calibration technique with a relative energy calibration technique
  such as photoelectron energy spectroscopy will be necessary to realize its
  full potential of achieving uncertainties as low as 1--10\,meV.}

\maketitle

\section{Introduction}
\label{sec:intro}

High-resolution astrophysical X-ray spectroscopy has become routine in
the last 20 years, with diffraction grating spectrometers on
\textit{Chandra} and \textit{XMM-Newton} providing resolving powers of
$\Delta \lambda/\lambda \sim 1000$
\citep{brinkman2000a,denherder:2001,canizares:2005,devries:2015}.
These instruments have enabled the measurements of the conditions in
the emitting plasmas, e.g., through observations of the triplets from
He-like ions, precision Doppler velocity and line shape measurements
in a variety of astrophysical plasmas, including stellar coronae and
winds, cataclysmic variables, X-ray binaries containing neutron stars
and black holes, supernova remnants, or outflows in active galactic
nuclei
\citep[e.g.,][]{paerels:2003,steenbrugge:2005,ishibashi:2006,miskovicova:2016,
  2014Sci...345...64K,drake:2019,nowak:2019}. Due to the success of
these measurements, future astrophysical X-ray observatories such as
\textsl{XRISM}, \textsl{Athena}, \textsl{Arcus}, or \textsl{Lynx},
envision spectral resolving powers as high as 5000, implying the
ability to accurately determine centroids to 10\,ppm, or
$3\,\mathrm{km}\,\mathrm{s}^{-1}$ absolute Doppler velocity
\citep{mcentaffer:2013,Heilmann2017, Barret2018, Bandler2019,
  Tashiro2020, Smith2020}. These instruments will open up the field of
spatially resolved, high-resolution X-ray spectroscopy, and will allow
scientists to access techniques that are currently not available to
X-ray astronomy such as X-ray Fine Structure Absorption measurements
for solids \citep{lee:2009a}, the imaging of velocity fields in galaxy
clusters \citep{cucchetti:2018}, or diagnosing the properties of the
Warm and Hot Intergalactic Medium \citep{walsh:2020}.

The ground and on-orbit calibration of existing and future instruments
as well as the interpretation of the existing and future observations
require accurately calibrated atomic transition energies
\citep[e.g.,][]{devries:2015,plucinsky:2017}. In one- and two-electron
ions, these energies are calculable with part per million (ppm)
accuracy for the astrophysically relevant atomic numbers less than 30
\citep[e.g.,][]{JohnsonSoff1985, Drake1988, Artemyev2005,
  yerokhin2019},
and theory has been experimentally benchmarked with precision as good
as 10\,ppm \citep[e.g.,][]{Kubicek2014,beiersdorfer2015}.

Inner shell transition energies in less-ionized species, neutral
atoms, molecules, and solids, are far more challenging to calculate
accurately, and thus must be obtained experimentally. These
experiments, however, rely on existing soft X-ray calibration
standards, which have limitations to their accuracy. We recently found
a discrepancy in the extensively used standard of the Rydberg
transitions of molecular oxygen of almost 0.5\,eV
\citep{leutenegger2020}, thus resolving a tension between
astrophysical and laboratory measurements of transitions of atomic
oxygen \citep{McLaughlin2013}, which had been calibrated against this
molecular standard \citep{WightBrion1974_O2}. Such
discrepancies raise the question of whether other commonly-used soft
X-ray standards may have errors of comparable magnitude, given that
many such standards are based on similar experimental techniques using
electron energy loss spectroscopy (EELS).

Even if the error in the earlier molecular oxygen
standard is an outlier, the typical experimental precision of soft
X-ray standards obtained with EELS is still of order 0.1\,eV (or
100\,ppm at 1\,keV), which is far too large to fully exploit the
capabilities of current and future X-ray astronomical and ground based
facilities, and not precise enough for the calibration needs of many
future instruments. Modern synchrotron facilities are capable of
sufficient photon fluxes and resolving powers that determining
centroids of peaks with statistical precision of 1--10\,ppm is routine in
a variety of experimental disciplines \citep[e.g.,][]{mueller2018,epp2015},
so to the extent that scientific results depend on the absolute
transition energies, calibration will often be the limiting factor.

The anticipated high precision of line energy measurements enabled by
high spectral resolution coupled with large photon fluxes in future
space-based observatories, as well as in high-performance synchrotron
beamlines, implies a need to reevaluate soft X-ray transition energies
of common elements and materials that have been used for energy
calibration using the same accurate standards used by
\cite{leutenegger2020}: highly charged ions (HCI) with one or two electrons.
To further illustrate the capabilities of these methods, in this paper
we present measurements of photoion yield spectra for
\CO around the oxygen K-edge, \SF around the F K-edge, and Ne around
its K-edge. These are calibrated using K-shell transitions of He-like
N, O, and F, respectively. The remainder of this paper is structured
as follows: In Sect.~\ref{sec:setup} we describe our experimental
setup, which combines a synchrotron beamline with an electron beam ion
trap (EBIT) to generate the calibrating ions and a gas cell, and
discuss the energy calibration and systematic limitations from this
setup. In Sect.~\ref{sec:results} we present the results of our
calibration of the photoionization spectra for neon, \SF, and \CO. In
order to understand the structure of the molecular edges in greater
detail, in Sect.~\ref{sec:theory} we then compare the experimental
results for the molecules with theoretical simulations. We summarize
the paper in Sect.~\ref{sec:conclusions}.

\section{Experimental setup and data analysis}\label{sec:setup}
\subsection{Experimental Setup}\label{sec:experiment}

\begin{figure*}
    \resizebox{\hsize}{!}{\includegraphics[width=\hsize]{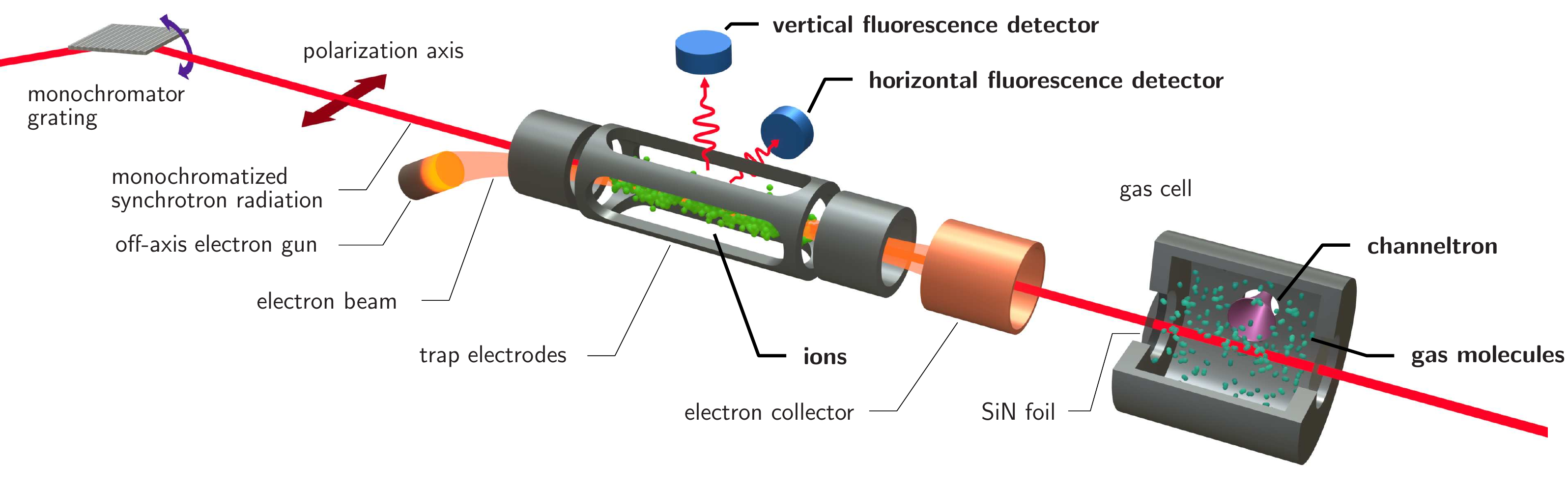}}
    \caption{Our scheme for simultaneous measurement of neutral gas
      photoionization and HCI fluorescence
      \citep[adapted from][]{leutenegger2020}. Monochromatic linearly
      polarized X-rays produced by the synchrotron beamline
      U49-2/PGM-1 enter the PolarX-EBIT endstation from the left,
      and excite the HCI. Subsequently, the fluorescence is detected by
      two silicon drift detectors. The off-axis electron gun allows the
      synchrotron X-ray beam to pass through to our second endstation,
      a low-pressure gas cell using a channeltron for detection of
      photoions.}
      \label{fig:experimental_setup}
\end{figure*}

Our experimental setup is depicted in
Fig.~\ref{fig:experimental_setup}. Monochromatic X-rays from a
synchrotron beamline pass through an EBIT, where they interact with
HCI. Fluorescence emission from these ions provides the basis for the
absolute calibration of the monochromator energy scale in our
experiment. The synchrotron radiation passes through the low density
plasma in the EBIT with virtually no attenuation, and then enters a
gas photoionization cell containing the atoms or molecules under
investigation. A channeltron inside the gas cell detects the ion yield
due to the interaction of the X-ray beam with the gas. The gas cell with
injected gases was operated with a pressure of few $10^-7$\,Torr.

Our setup used the PolarX-EBIT \citep{micke2018}, which features an
off-axis electron gun, enabling the photon beam to pass through the
EBIT. The electron beam is tuned to an energy sufficent to ionize
atoms entering the trap up to the He-like charge state, but staying
below the threshold for K-shell excitations, and also avoiding
dielectronic-recombination resonances. The X-ray photons interacting
with the ions thus produce a K-shell fluorescence signal that is
uncontaminated by X-rays following collisional excitation. We measured
this fluorescence signal with silicon-drift detectors (SDDs) that are
mounted perpendicular to the electron beam axis.

For H-like and He-like systems it is possible to calculate the
transition energies with uncertainties of $\lesssim$1\,meV
\citep{yerokhin2019}. This \emph{ab initio} provides the absolute
calibration reference for our measurements. Since our experiment
allows us to measure the fluorescence in the EBIT simultaneously with
the ion yield in the gas cell, we avoid problems that are intrinsic
to non-simultaneous energy calibrations.

For \NE, \CO and \SF investigation, we measure the fluorescence of
He-like fluorine, nitrogen, and oxygen, respectively. We induce it
with soft X-ray photons provided by the BESSY II plane-grating
monochromator (PGM) beamline U49-2/PGM-1 \citep{sawhney2001}. Because
of the linear polarization of the beam and the dipolar character of
the resonances, there is a strong dependency of the fluorescence on
the viewing angle \citep{bernitt2013}. Therefore, we used two SDDs,
one aligned parallel to and the other perpendicular to the
polarization axis. Polarization also slightly affects the ion-yield
measured in the gas cell. The channeltron was aligned parallel to the
polarization axis, but because it was close to the photon beam, it has
a finite angular acceptance. The acquired photoion spectra showed
features excited by both polarization axes, albeit with a stronger
contribution from the parallel axis.

Individual scans for each of the three gases were performed in
equidistant energy steps from low to high energies, scanning the
photon energy in ranges of 866--871\,eV for \NE, 533--540\,eV for \CO,
and 684--705\,eV for \SF. At each scan step, the integrated
ion-production rate and HCI-fluorescence rates were recorded together
with the nominal energy of the beam line. To achieve the highest
possible accuracy and minimize uncertainty, the calibration line must
lie within the scan range. This was possible for the \CO and \SF
scans. For the \NE scan, the chosen calibration line was 10\,eV lower
in energy and had to be recorded in a separate scan. The details of
data recording and processing are described in \citep{leutenegger2020},
where the same setup was used.

\subsection{Energy Calibration}\label{sec:calibration}

The nominal calibration of the beamline wavelength scale uses the
grating equation
\begin{equation}
  m N \lambda = \cos\alpha - \cos\beta
\end{equation}
where $N$ is the line density of the grating and $m$ the diffraction
order. In our experiment, $m=1$ and $N = 602.4\,\mathrm{mm}^{-1}$. The
incident and reflection angles $\alpha$ and $\beta$ are measured with
respect to the plane of reflection. These angles are determined from
the rotation angles of the mirror and grating using high-precision
rotation encoders. Typically, the true wavelength of the beamline has a
slight offset from the nominal value derived using the encoder
positions. This offset can be corrected using the calibration lines.
In many experiments it is common practice to apply a linear offset to
wavelength or energy based on a calibration feature. However, since
the grating equation is nonlinear, this introduces a systematic error
that increases with separation from the calibration feature.
Specifically, in energy space the grating equation is given by
\begin{equation}\label{eq:grating}
  E = \frac{h c m N}{\cos\alpha - \cos\beta}
\end{equation}
where $h$ is Planck's constant and $c$ is the speed of light. We used
the defined CODATA 2018 value $hc = 1\,239.841\,984\,\mathrm{eV}\,\mathrm{nm}$
\citep{newell2018,bipm2018}\footnote{The nominal energies reported by
  the beamline use $hc = 1\,239.86\,\mathrm{eV}\,\mathrm{nm}$.}.

The angles comprise two parameters while the selection of energy fixes
only one degree of freedom. The remaining degree of freedom is fixed
by $\alpha$ and $\beta$ adhering to the fixed-focus condition
\citep[equation 2, converted to our angle convention]{follath2001}
\begin{equation}
  \sin\beta = c_\textnormal{ff} \sin\alpha,
\end{equation}
with $c_\textnormal{ff}$ set to 2.25 for U49-2/PGM-1. This fixed focus
condition ensures that the image of a source at a fixed distance to the
grating is projected to a fixed point behind the grating with a scaling,
$c_\textnormal{ff}$, that is independent of the energy.

Throughout our campaign we found a discrepancy of more than 3\,eV at
the energy of the $\mathrm{O}^{6+}\,\rm 1s^2\,{}^1S_0 \rightarrow 1s2p\,{}^1P_1$
transition. We will call this line O$_\textnormal{w}$ in the following
\citep{gabriel1972}; its theoretical energy is 573.961\,eV
\citep{Drake1988}.
Assuming that the accuracy of the angle steps is stable (at least over
single scans, containing 100--1000 steps each), this discrepancy must
be due to an offset in the angles such that $\alpha = \alpha' + \Delta\alpha$,
$\beta = \beta' + \Delta\beta$, where $\alpha'$ and $\beta'$ are the
incident and reflected angles of the photons as reported by the
beamline.

A single calibration feature only permits to determine either
$\Delta\alpha$ or $\Delta\beta$. A natural choice for their relation
is to have the corrected values fulfill the fixed-focus condition,
which can be approximately achieved through
\begin{equation}
  \Delta\alpha c_\textnormal{ff} = \frac{\cos\beta'}{\cos\alpha'}\Delta\beta.
\end{equation}
Although we emphasize that this choice is not {\it a priori}
theoretically motivated, we found that the derived energy scale is not
sensitive to a particular relation between $\Delta\alpha$ and
$\Delta\beta$ if both are sufficiently small, and the calibration is
applied to an appropriately small energy range. The reason is that
small changes of $\alpha$ or $\beta$ have the same effect, that is,
shifting the energy scale. It is only for large changes of $\alpha$
or $\beta$ that the slope of the calibration changes.

The final calibration of the energy of the gas cell measurements is
achieved through a simultaneous fit to both the photoion yield
spectrum in the gas cell and the fluorescence spectrum in the EBIT.
Using theoretical values for the energies of the fluorescence lines,
the free parameters of the fit are the angular shifts of
$\alpha$ and $\beta$, the energies of the photoionization resonances
in the gas cell spectrum, and their respective widths. Each
fluorescence line and photoionization resonance is modeled with a
Voigt profile, with the Gaussian $\sigma$ and Lorentzian $\Gamma$
parameters representing a combination of instrument profile, natural
linewidth, and thermal Doppler broadening; both spectra also include
background which we model with a energy independent constant. In the
fluorescence spectrum, it is mainly caused by the
high-energy tail of the pulse height distribution of low-energy
photons detected by the SDDs. In the photoionization spectra, it
results from residual gases that do not have resonant features in the
bands of interest, and can thus be treated as a constant contribution
to the detected signal.

All of the fits are evaluated using the Cash statistic
\citep{cash1979}, a version of the likelihood ratio test that is
appropriate for Poisson distributed data.
Since we model the calibration and ionization data simultaneously, it
is possible to estimate confidence intervals for our parameters of
interest by confidence search \citep{cash1976}. These intervals
describe the total statistical uncertainty for each emission line in
the photoionization spectrum, including the one from the calibration
measurement (but excluding systematic uncertainties, as discussed
below). The confidence intervals derived using this approach cover the
90\% uncertainty interval and are typically in the range between 1 to
10\,meV.

\subsection{Systematic limitations}\label{sec:limits}

\begin{figure}
    \resizebox{\hsize}{!}{\includegraphics{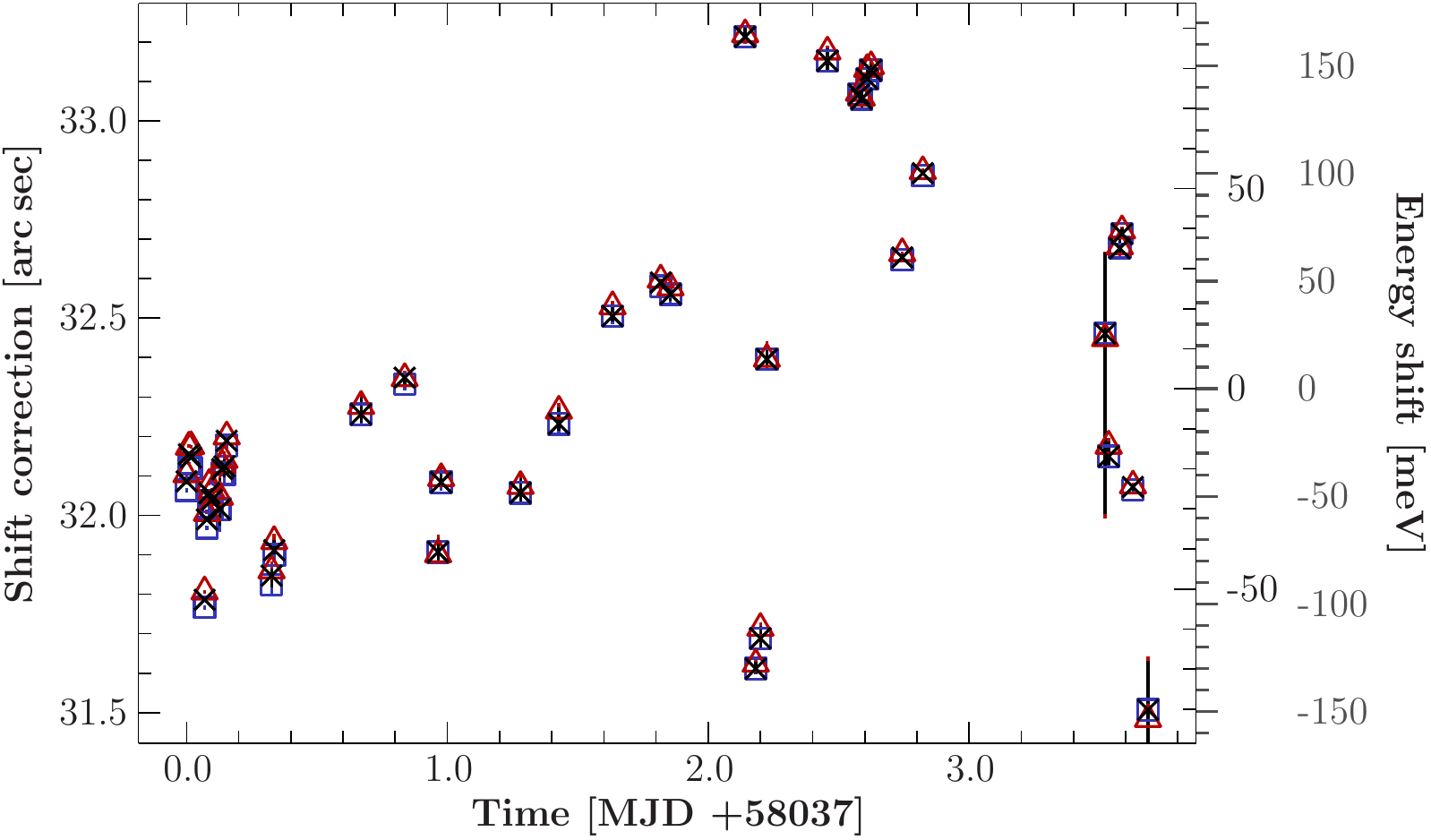}}
    \caption{Measured position of the O\textsubscript{w} resonance in
      terms of the angle correction at different times during the
      campaign. Black crosses and red triangles respectively indicate
      the reported energy after and before the photon measurement, and
      blue squares their mean (see text for details). The right scale
      shows the translated energy spread around the mean of all
      measurements at the O$_\textnormal{w}$ line (inner) and the Ne
      1s--3p transition (outer). Measurements of this line were mainly
      used to benchmark the X-ray beam at different settings causing a variation in the uncertainty.}
      \label{fig:long-variability}
\end{figure}

The calibration uncertainty of the photoionization spectra is
dominated by systematic terms. As discussed in Sect.~\ref{sec:setup}
and~\ref{sec:calibration}, the statistical uncertainties on the
calibration are typically smaller than 10\,meV, while the theoretical
uncertainties in our calibration line energies are smaller than
1\,meV.

In our current setup, a contribution to the uncertainty much larger
than those comes from the stability of the beamline. We can estimate
the long-term variability of the beamline from scans measuring the
same transition throughout the measurement campaign. The fluctuation
of the shift parameter, $\Delta\beta$, is shown in
Fig.~\ref{fig:long-variability} for repeated measurements of the
O\textsubscript{w}. On the right-hand $y$-axis, we display the
corresponding effect of such an angular shift on the energy
calibration at the O\textsubscript{w} line (574\,eV) and the neon K-edge (870\,eV).
In our experiment we requested the reported monochromator energy and
angle settings twice for each scan step; once before data acquisition
with the SDD and once after. We found that the reported energy values
before the SDD acquisition often showed unreasonably high fluctuations,
probably attributable to the relaxation of the beamline to the selected
energy immediately after moving the monochromator, even after the
allowed settling time. For this reason we only used the values reported
after each scan step for our further analysis.
As more extensively discussed  in the supplemental
material of \citep{leutenegger2020}, based on repeated scans of
multiple closely-spaced photoionization lines in the gas cell, and
also on studies of the shapes of single fluorescence lines in
the SDD, we conclude that such large shifts do not occur in single
scans; however, shifts of up to 40\,meV can be expected near
O\textsubscript{w}. Given that the energy shifts for a fixed angular
shift become larger at higher energies, we estimate that the
systematic energy shift at the Ne K-edge can be as high as 100\,meV;
we discuss this further in Sect.~\ref{subsec:neon}.

\section{Energy calibration of \NE, \CO, and \SF}\label{sec:results}

We now discuss the results of modeling each of the photoionization
spectra measured for neon, \SF and \CO.

\subsection{\NE Rydberg series}\label{subsec:neon}
We calibrated our scan of atomic neon using the F$^{7+}\,K_\beta$
transition
\citep[$E_{\textnormal{F}\,K_\beta}$ = 857.5108(7)\,eV,][]{yerokhin2019}. 

This line was scanned before and after the actual ionization
measurement of neon and, therefore, not simultaneously. As discussed
above, this adds an additional uncertainty which can, in principle, be
as large as 150\,meV (Fig.~\ref{fig:long-variability}). The angular
shift corrections measured for the two F$^{7+}\,K_\beta$ scans differ
by $0.2''$, corresponding to about 30\,meV at
$E_{\textnormal{F}\,K_\beta}$. Instead of using the averaged shift
correction as obtained from both calibrations, we weigh
the shift correction of the neon data with Student's~$t$ distribution
\citep{student1908}. In this way we can estimate the statistical
uncertainty due to the variation of the calibration by assuming that
these are drawn from a normal distribution. Just accounting for
statistical variations, the resulting 90\% confidence interval for the
energy of the neon lines is $\pm15$\,meV. The systematic uncertainty
can not be quantified directly but can be deduced by comparison to
previous experiments. Overall we estimate a 100\,meV calibration
uncertainty.

\begin{figure}
    \resizebox{\hsize}{!}{\includegraphics{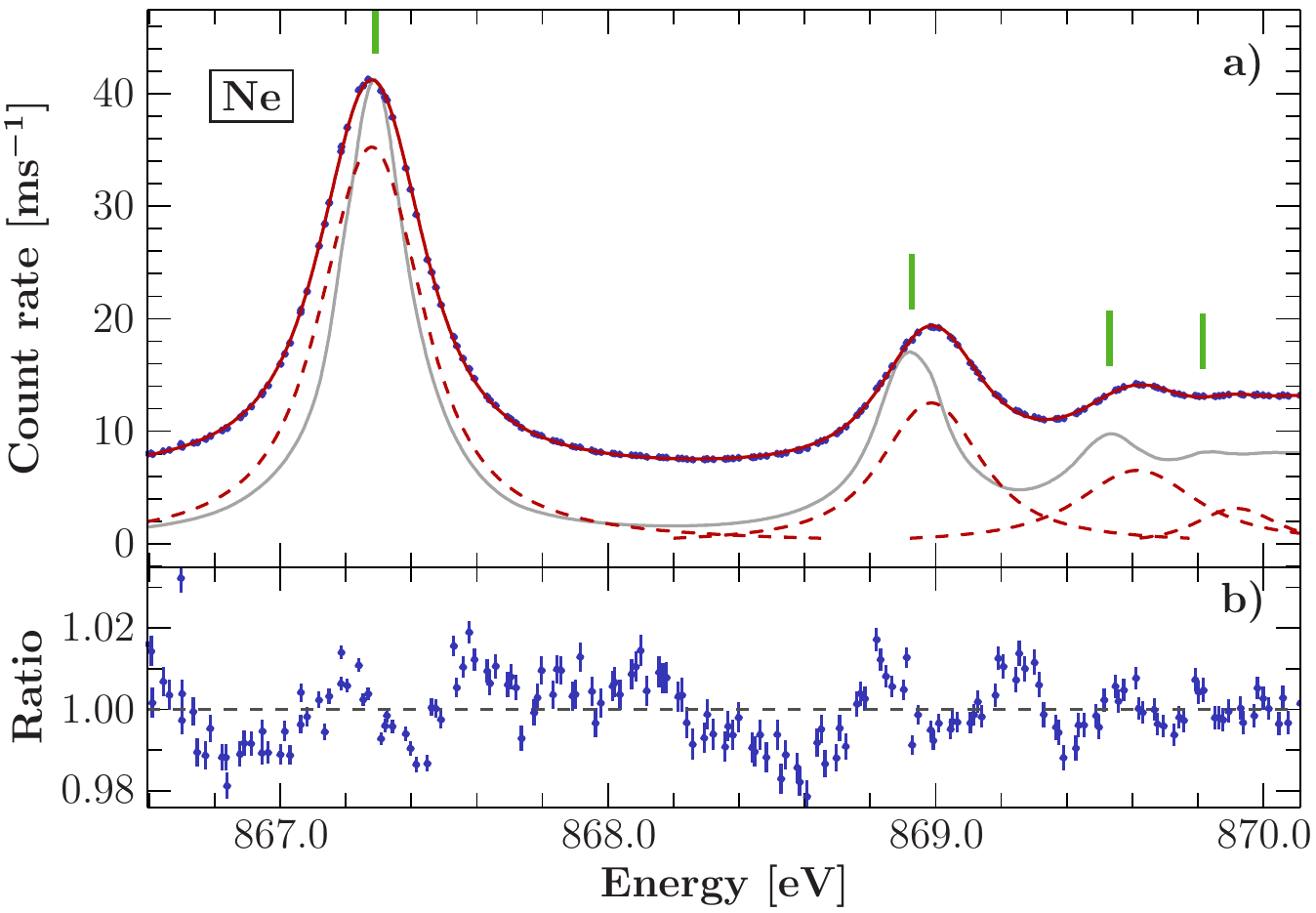}}
    \caption{\textbf{a:} Neon spectrum (blue points) calibrated using measurements
      of F\,$K_\beta$ in scans before and after the photoionization
      measurements. The neon emission lines are modeled by a sequence
      of Voigt profiles to determine the peak positions (red solid
      line, components red dashed lines). The green vertical bars
      indicate the line positions as reported in \citep{mueller2017} and
      the gray solid line outlines their data (scaled to the 1s--3p
      transition). \textbf{b:} The ratio between the data
      and the model. The calibration line was modeled with Voigt
      parameters $\sigma = 0.172$\,eV and $\Gamma < 0.001$\,eV.}
      \label{fig:neon}
\end{figure}

The Rydberg series (Fig.~\ref{fig:neon}) is modeled by a set of five
Voigt profiles without constraints on the line shape parameters. The
scan range does not reach up to the series limit such that it is not
possible to include a component for the edge without constraints on
its location. The model used by \cite{mueller2018}, where the
positions are constrained by a Rydberg series modified by a quantum
defect, is not describing our data to a satisfying level. Therefore we
did not constrain the line positions and we also did not include a component
for the ionization edge.
Ignoring contributions of the ionization edge to the high energy part
of the scan causes the fifth Voigt profile to model all contributions
from higher Rydberg transitions and the ionization edge. This behavior of the model can
have an effect on the position of the 1s-6p line, but we expect that
the effect on the lower transitions is only marginal and below the
uncertainty.
The resulting model is shown in Fig.~\ref{fig:neon} and has only a few
residual patterns left. A part of these residuals can be attributed
to the uncertainty (or jitter) of the reconstructed energy grid.
We verify this by fitting the same model to the data using the
nominal energy grid. Here the residuals cluster around the wings of
the model lines since a jitter in the energy grid has a larger impact
at energies the derivative of the model has a larger absolute value. For
the reconstructed energy grid these residual patterns are stretched
over a wider energy range.

\begin{table}
    \def\arraystretch{1.5}
    \caption{Measured \NE Rydberg transitions lines
    calibrated against the F\,$K_\beta$ line.}
    \label{tab:neon}
    \begin{tabular}{lcc}
        \hline\hline
                   & \multicolumn{2}{c}{Energy [eV]} \\
        Transition & This work\tablefootmark{(a)} & M\"uller et al.\citep{mueller2017}\tablefootmark{(b)} \\
        \hline
        1s--3p & {\NeonI} & 867.290 \\
        1s--4p & {\NeonII}  & 868.928 \\
        1s--5p & {\NeonIII}  & 869.530 \\
        1s--6p & {\NeonIV}  & 869.815 \\
        \hline
    \end{tabular}\\
    \tablefoottext{a}{Calibrated against F\,$K_\beta$ \citep[857.5108(7)\,eV,][]{yerokhin2019}}\\
    \tablefoottext{b}{Calibrated against neon 1s--3p \citep{wuilleumier1971}}
    \tablefoot{Statistical
    uncertainties of the peak positions are $\pm15$\,meV but are largely
    exceeded by systematic variations of up to 100\,meV (see text). Recent high
    resolution measurements of the neon Rydberg series are given for comparison \citep{mueller2017}.}
\end{table}

Our determined line positions are given in Table~\ref{tab:neon}. We
compare these results with those found in \citep{mueller2017}, which
have been calibrated using the measured 1s--3p transition of
\citep{wuilleumier1971}. The agreement of the first line is very good,
while the subsequent lines diverge more with higher energy.
The difference of order 50-100 meV in the higher-$n$ lines is
consistent with the amplitude of drift observed in the energy scale
of U49-2/PGM-1, as discussed in the supplemental material of
\citep{leutenegger2020}.

\subsection{\CO Oxygen K-edge}\label{subsec:co2}
\begin{figure}
  \resizebox{\hsize}{!}{\includegraphics{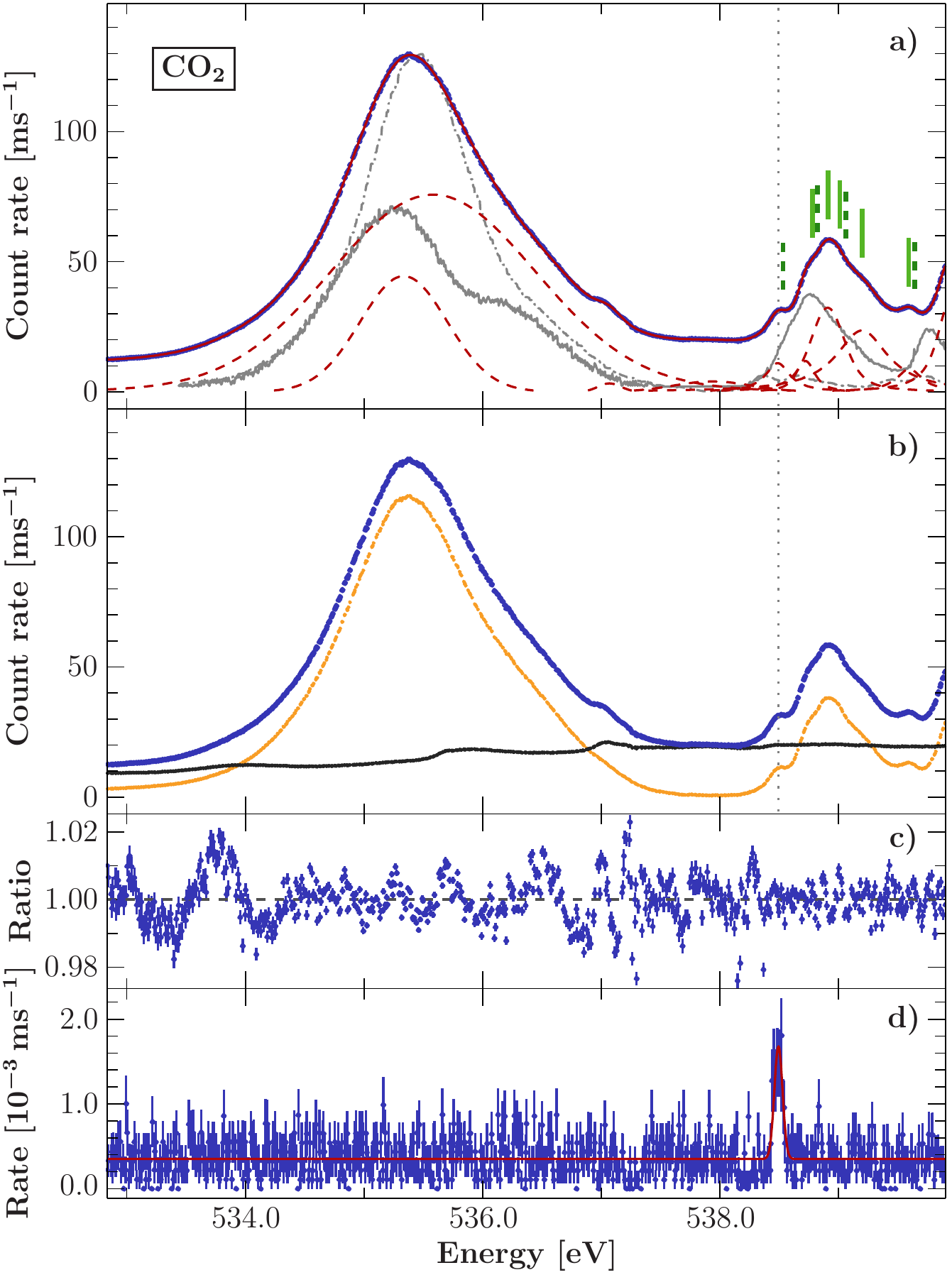}}
  \caption{\textbf{a:} Calibrated \CO spectrum (blue points) based
    on simultaneous measurements of N\,$K_\epsilon$. Emission lines are
    modeled by Voigt profiles (red solid line, components red dashed lines).
    Model components may represent multiple unresolved transitions. The
    green lines indicate the transition energies in the Rydberg
    complex reported in \citep{adachi2005} for the two resolved
    symmetry directions 0$^\circ$ (solid) and 90$^\circ$ (dashed).
    Solid gray and dashed-dotted gray line indicated their data, for
    0$^\circ$ and 90$^\circ$, respectively. The dotted gray vertical
    line indicates the location of the calibration line for our data.
    \textbf{b:} Residual water vapor in the gas cell adds additional spectral features.
    The background was estimated from data from a second gas cell (black points);
    The corrected spectrum (orange points) shows the difference between
    the data of the two gas cells. The uncorrected data is again given here
    for reference (blue points).
    \textbf{c:} Residuals between model and data as
    ratio.
    \textbf{d:} Sum of the fluorescence spectra produced in
    the EBIT measured by the two SDDs. The calibration line was modeled with
    Voigt parameters $\sigma = 0.082$\,eV and $\Gamma < 0.001$\,eV.}
    \label{fig:co2}
\end{figure}

We measured the photoionization yield of \CO in the range 533 to
540\,eV. The calibration of the energy grid is based on the
theoretical predictions of the He-like nitrogen transition
N\,$K_\epsilon$
\citep[$E_{\textnormal{N}\,K_\epsilon} = 538.4924(3)$\,eV,][]{yerokhin2019}.

This line was measured simultaneously with the ionization spectrum, thus
significantly reducing the overall uncertainty. The measured \CO
spectrum is rich, featuring (in our spectrum) unresolved vibrational
structure, and showing a mixture of lines from both polarizations
\citep{adachi2005,okada2002}. To determine the transition energies, we
empirically modeled the spectrum with a set of 10 Voigt profiles, which
in many cases represent blends of unresolved emission lines. The choice
of 10 lines is only supported by the number of features which are
identifiable by eye. The background is modeled with an energy
independent constant.
Fig.~\ref{fig:co2} shows the calibrated data and
best fit model. This best fit model is reasonably good, with
residuals comparable to the neon measurement. On
closer inspection, the spectrum shows a rich structure which is only
barely resolved in our data but clearly visible in recent resonant
inelastic X-ray scattering (RIXS) measurements \citep{soderstrom2020}.

This large number of parameters in the empirical model poses a
difficult problem for classical fit algorithms, especially with the
addition of the calibration function itself. To find the minimum of
this function, we made use of the Markov Chain Monte Carlo (MCMC)
algorithm proposed in \citep{foreman2013}. It explores the
probabilistic parameter space, and additionally gives the parameter
uncertainty. The resulting 90\% confidence for the first 9 line
profiles is $\pm3$\,meV. The tenth line is only partly covered by
the scan, and therefore not well constrained.

\begin{table}
  \caption{\CO measured transitions in our calibration.}
  \label{tab:co2}
    \def\littleskip{16pt}
    \begin{tabular}{lccc}
      \hline \hline
                 & \multicolumn{2}{c}{Energy [eV]} \\
      Transition & This work\tablefootmark{(a)} & Okada \citep{okada2002}\tablefootmark{(b)} &
        Adachi \citep{adachi2005} \\
      \hline
      \tboxr{$\pi^*$\\3s$\sigma$}         & \tboxlrn{\COI\\\COII}  & 535.4\makebox[0pt][l]{\tablefootmark{(c)}} & 535.4\makebox[0pt][l]{\tablefootmark{(c)}} \\
      \tboxrn{contam.}                    & \tboxlrn{\COIII}       & \samebox{000.0}{---}                       & \samebox{000.0}{---} \\
      \tboxrn{contam.}                    & \tboxlrn{\COIV}        & \samebox{000.0}{---}                       & \samebox{000.0}{---} \\
      \tboxrn{3p$\pi_u$}                  & \tboxlrn{\COV}         & \tboxln{538.53}                           & \tboxln{538.53} \\
      \tboxr{3p$\sigma_u$\\3p$\pi_u$}     & \tboxlrn{\COVI}        & \tboxl{538.78\\538.83}                     & \tboxl{538.78\\538.82} \\
      \tboxr{4s$\sigma_g$\\3p$\sigma_u$}  & \tboxlrn{\COVII}       & \tboxl{538.93\\539.04}                     & \tboxl{538.91\\539.06} \\
      \tboxr{4s$\sigma_g$\\3p$\sigma_u$}  & \tboxlrn{\COVIII}      & \tboxl{539.18\\539.30}                     & \tboxln{539.20} \\
      \tboxrn{3d$\pi_g$}                  & \tboxlrn{\COIX}        & \tboxln{539.67}                           & \tboxln{539.64} \\
      \hline
  \end{tabular}
  \tablefoottext{a}{Calibrated against N\,$K_\epsilon$ \citep[$538.4924(3)$\,eV,][]{yerokhin2019}.}\\
  \tablefoottext{b}{calibrated against \CO transitions from \citep{prince1999}.}\\
  \tablefoottext{c}{Unresolved blend of $\pi^*$ and 3s$\sigma$, reported in \citep{wight1974}.}
  \tablefoot{For comparison the experimental values of \citep{okada2002,adachi2005}
    are listed. Assignments are based on the assignments of \citep{okada2002}. Line blending
    and mixing is indicated by braces. We estimate the uncertainty of our energy scale to
    40\,meV (see text).}
\end{table}

We list the resulting line positions in Table~\ref{tab:co2}, where
the assignments are by strongest contribution to our model based on the
measurement of \citep{okada2002}. The resonance peak shows two main features
\citep[e.g.,][]{adachi2005} generally associated to the valence
orbital and contribution from the 3s$\sigma$ state. A small emission
line is visible at the shoulder of the resonance peak together with an
excess of events between the resonance and the Rydberg complex
compared to recent high resolution RIXS measurements
\citep{soderstrom2020}. This excess can be attributed to residual
water vapor in the gas cell.

An estimate of the residual gas is obtained from a second gas cell
operated upstream of the first cell and separated from it by a thin SiN
window. The second cell was operated with no sample gas injection, and
therefore all photoions detected are from background gases. In
principle, the background gas composition in the two cells may be
different. However, due to insufficient bakeout, the residual gas in
our vacuum chambers was dominated by water vapor, as can be seen by
comparing the features in the background gas spectrum to previously
published measurements of water vapor \citep{WIGHT1974_H2O_etal}. As
indicated in Fig.~\ref{fig:co2}, we see that the residual gas spectrum
explains the feature on the high energy side of the $\pi^*$ resonance
as well as the unexpectedly high amplitude of the continuum between the
resonance and the Rydberg complex. Because we could not be certain that
the amplitude of the background spectrum was the same in both cells, we
cannot use the second cell to correct the first. However, we can try
modeling the background in the first cell based on the spectrum of the
second and assess the impact on our fit results. We found that the
energy of the $3s\sigma$ peak moved to slightly higher energy, while
the other peak energies were unaffected.
We attribute the remaining residuals in the  $\pi^*$ resonance to a
combination of unresolved vibrational structure
\citep{ljungberg2017,vazdacruz2021} and a possible non-ideal instrument lineshape.
The dominant uncertainty in the transition energy determination is drift in
the monochromator energy scale. Based on the analysis in
\citep{leutenegger2020} we estimate this uncertainty to be 40\,meV for these
lines.

Direct comparison of the results with the literature is in general not
possible due to the blending of transitions. However, the 3p$\pi_u$
transition is easily identifiable in our scan as well as in recent
measurements \citep{soderstrom2020,adachi2005,okada2002}.
Additionally, its energy is very close to our calibration line and,
hence it has much smaller shifts due to drift.
Using this as a reliable reference energy, we see that the Rydberg
complex from \citep{okada2002} appears at slightly higher energies.
The result of \citep{okada2002}, calibrated using \CO measurements from
\citep{prince1999}, which in turn are calibrated against 0$_2$ measuremnts from \citep{hitchcock1980}\footnote{For
  the calibration with respect to O\textsubscript{2}, \citep{prince1999} references
  \citep{sodhi1984}. However, the given value is actually obtained from \citep{hitchcock1980}.}.
Given the uncertainties of 100-200\,meV in their work, we conclude that our
measurements of the transition energy of
3p$\pi_u$ agree.
Similar EELS measurements \citep{eustatiu2000b} also place the
$\pi^*$ resonance at a higher energy, but comparison of the Rydberg complex is
difficult due to their limited energy resolution and lack of polarization selectivity.

\subsection{\SF Fluorine K-edge} \label{subsec:sf6}

We scanned the fluorine K-edge of \SF in the range from 685 to 705\,eV
measuring the photoionization yield of the gas in the gas cell. The
calibration is based on the O\,$K_\gamma$ transition of
He-like oxygen
\citep[$E_{\textnormal{O\,K}_\gamma} =
697.7859(5)$\,eV,][]{yerokhin2019}.

\begin{figure}
  \resizebox{\hsize}{!}{\includegraphics{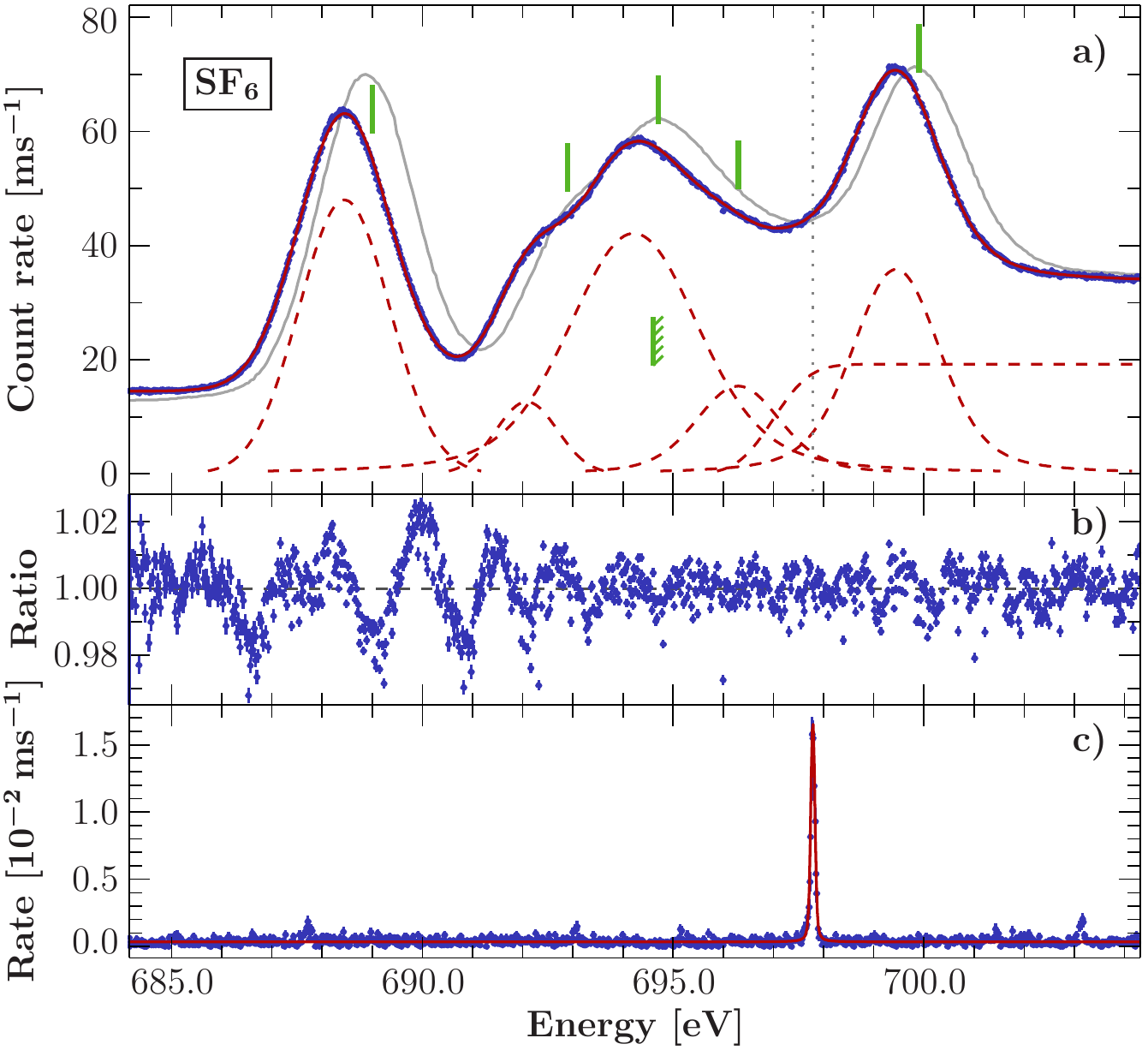}}
  \caption{\textbf{a:} \SF photionization spectrum (blue points) calibrated by
    simultaneous measurement of the O\,$K_\gamma$ transition and
    modeled by five Voigt profiles and one error function (red solid
    line, components red dashed lines). The position of the
    calibration line is indicated by the dotted vertical line. The
    solid green lines indicate the measured peak positions from
    \citep{hudson1993} and the edge as measured in
    \citep{siegbahn1969} (indicated with diagonal marks). The gray
    curve outlines the measurements of \citep{hudson1993} scaled
    to match the present results.
    \textbf{b:} Ratio between the data and the best-fit model.
    \textbf{c:} Sum of fluorescence spectra measured with
    the two SDDs. The calibration line is modeled with the Voigt parameters
    $\sigma = 0.079$\,eV and $\Gamma = 0.031$\,eV}
    \label{fig:sf6}
\end{figure}

Following \citep{hudson1993}, we describe the spectrum by a sequence
of five Voigt profiles together with an error function to account for the
photoelectric absorption edge at the Rydberg series limit. In
\citep{hudson1993} an arctangent function convolved with a Gaussian was
used to describe the edge, but given our resolution we cannot discriminate
between these choices. Hence, for simplicity we used only an error
function to model the edge. The calibrated data and resulting best fit are displayed in
Fig.~\ref{fig:sf6}. Similarly to \citep{hudson1993}, we have to add a line
around 696\,eV, otherwise an excess of events remains visible in the
data in comparison with the model. Further justification for emission at this energy is given by
theoretical predictions (see Sect.~\ref{sec:sf6-theo}). From the
confidence calculations, we estimate the 90\% uncertainty to
$\pm3$\,meV for the line positions; the edge energy has an uncertainty of
$+2$\,meV. The systematic uncertainty is dominated by drift in the beamline
energy scale, and based on \citep{leutenegger2020}, we estimate it to be
60\,meV.

\begin{table*}
  \caption{Calibrated \SF transitions and comparison to other publications.}
  \label{tab:sf6}
  \begin{tabular}{lccccccc}
    \hline\hline
               & \multicolumn{7}{c}{Energy [eV]} \\
               \multicolumn{3}{c}{This work} && \multicolumn{4}{c}{Other experiments} \\ \cline{1-3} \cline{5-8}
    Symmetry & Experiment\tablefootmark{(a)} & Theory && Eustatiu\tablefootmark{(b)} & Francis\tablefootmark{(c)} &
    Hudson\tablefootmark{(d)} & Hitchcock\tablefootmark{(e)} \\
    \hline
    \tboxrn{a\textsubscript{1g}}                                 & \SFI    & \samebox{000.00}{---}& & 687.9 & 688.0 & 689.0 & 688.0 \\
    \tboxr{\underline{a\textsubscript{1g}}\\t\textsubscript{1u}} & \SFII   & \tboxl{\underline{691.59}\\692.23} && 691.4 & 692.4 & 692.9 & 692.6 \\
    \tboxr{t\textsubscript{1u}\\e\textsubscript{g}\\t\textsubscript{2g}\\\underline{t\textsubscript{1u}}}
                                                                 & \SFIII  & \tboxl{693.67\\693.81\\693.95\\\underline{694.19}} & & 693.5 & 694.0 & 694.7 & 694.6 \\
    \tboxr{t\textsubscript{1u}\\t\textsubscript{1u}\\\underline{t\textsubscript{2u}}}
                                                                 & \SFIV   & \tboxl{695.11\\695.43\\\underline{695.69}} & & \samebox{000.0}{---} & \samebox{000.0}{---} & 696.3 & \samebox{000.0}{---} \\
    \tboxrn{I.P.}                                                & \SFEdge & \samebox{000.00}{---} & & 694.6\makebox[0pt][l]{\tablefootmark{(f)}} & 694.6\makebox[0pt][l]{\tablefootmark{(f)}} & \samebox{000.0}{---} & 694.6\makebox[0pt][l]{\tablefootmark{(f)}} \\
    \tboxrn{t\textsubscript{2g}}                                 & \SFV    & \tboxln{699.51} & & 698.8 & 698.9 & \samebox{000.0}{699.9} & 699.1 \\
    \hline
  \end{tabular}
  \def\arraystretch{1}
  \halign{#\hfil\tabskip=12pt && #\hfil\crcr
    \tablefoottext{a}{\parbox{0.5\textwidth}{Calibrated against O\,$K_\gamma$  \citep[$697.7859(5)$\,eV,][]{yerokhin2019}.}} & \tablefoottext{b}{Eustatius et al. \citep{eustatiu2000a}.}\cr
    \tablefoottext{c}{Francis et al. \citep{francis1995}.} & \tablefoottext{d}{Hudson et al. \citep{hudson1993}.}\cr
    \tablefoottext{e}{Hitchcock \& Brion \citep{hitchcock1978}.} & \tablefoottext{f}{Determined from XPS \citep{siegbahn1969}.}\cr
  }
  \tablefoot{Theory values
    obtained by TDDFT calculations (see
    Sect.~\ref{sec:sf6-theo}. The first excitation
    (a$_\textnormal{1g}$) is chosen to align with the experiment.
    Assignments are based on these calculations, where the largest
    contribution to each spectral feature is underlined.
    Line blends are indicated by braces. For comparison, the results
    of selected publications are listed. We estimate the uncertainty
    of our energy scale to 60\,meV.}
\end{table*}

Overall, the empirical model describes the data very well.
However, the lowest energy line has significant residuals.
These residuals may originate from a combination of unresolved
vibrational structure \citep{ljungberg2017,vazdacruz2021} and a
possibly non-ideal instrument lineshape.
The spectrum has been measured several times in the past with varying
results \citep[e.g.,][]{siegbahn1969,gelius1974,hitchcock1978,hudson1993,francis1995,eustatiu2000a}.
Many of the EELS measurements use the measurements in
\citep{hitchcock1978} for calibration, which itself is based on
measurements from \citep{hitchcock1977}.
The EELS measurements have a difference of ${\sim}500\,\mathrm{meV}$
with our result, two to three times more than their claimed
uncertainty, but in agreement with the discrepancy in
\citep{leutenegger2020}. The photoionization measurements from
\citep{hudson1993} also show a shift to higher energies, but their
calibration was provided only by the used beamline. The results of
this work together with selected previous results are given in
Table~\ref{tab:sf6}. It is evident that the often used value of the
edge energy as reported in \citep{siegbahn1969} is not compatible with
our results (indicated in Fig.~\ref{fig:sf6}). A similar observation
can be made from the spectrum given in \citep{hudson1993}; however,
their edge location is not quantified.

\section{Modeling K-edge absorption spectra from first principles}\label{sec:theory}\label{sec:sf6-theo}\label{sec:co2-theo}

In order to assist in the interpretation of the experimental data we
performed \emph{ab initio} TDDFT simulations of the oxygen K-edge
excitations in \CO (shown in Fig.~\ref{fig:co2_simu}) and fluorine
K-edge in \SF (shown in Fig.~\ref{fig:sf6_simu}). The computation of
the molecular orbitals associated with the excited states allows us to
attach symmetry labels to the experimental peaks. Moreover, the
arbitrarily high resolution of the simulated spectra can help to
understand whether observed peaks originate from single broadened
transition lines or if there is a richer spectral structure which
cannot be resolved experimentally.

\subsection{Calculation details}
\label{subsec:theory_tddft}
For the simulation we employ Time-Dependent Density Functional Theory
(TDDFT) as implemented in the ORCA quantum chemistry code
\citep{ORCA_main}. We used a minimally augmented diffuse quadruple
zeta basis set \texttt{ma-def2-QZVPP}
\citep{Weigend2005_basis,Zheng2011_basis} and the Coulomb fitting
auxiliary basis \texttt{def2/J} in combination with the hybrid
functional \texttt{PBEh}$\alpha$ \citep{Perdew1996_functional}. All
calculations employed the RIJCOSX approximation
\citep{Neese2009_rijcosx}. Moreover, we considered only purely
electronic effects and neglected vibrational modes of the molecules.
The only free parameter of the DFT simulation is the mixing factor
$\alpha$ of the hybrid functional, for which we found the best
agreement with the present experiments at a value of $\alpha=35\,\%$. The
infinitely sharp transitions of the TDDFT simulation were subsequently
broadened (convolved) for comparison with experimental data.
Following common procedure, we employed an energy-independent Gaussian of 1.8\,eV (i.e.,
accounting for measurement effects) and an energy-dependent Lorentzian
(i.e., life time broadening) of the order of 0.10--4.47\,eV.
Since the energy offset of the TDDFT spectra
is known to be unreliable, following common practice the simulated
spectra were shifted to align the lowest lying excitation with the
experimental data.

\subsection{General remarks}
\label{subsec:theory_general}
While our TDDFT calculation includes core-hole effects
\citep{Petersilka1996,Bunau2012} beyond a simple mean-field limit,
excitonic multiplet splittings of the excited states are negligible in
K-edges (as opposed to, e.g., transition metal L- or rare-earth
M-edges). Therefore, a single-particle picture can be used to
interpret the excitations as the promotion of an oxygen or fluorine 1s
core electron into ``unoccupied'' molecular orbitals. We can exploit
this single-particle nature and associate to each peak in the spectrum
a corresponding single electron molecular-orbital computed from the
self-consistent DFT (and plotted with the Avogadro program
\citep{Avogadro2012}). Its symmetry (and degeneracy) then allows us to
classify the excitations in terms of irreducible representations of
the molecules point-group. For more details on approximation and
simulation strategies for (especially oxygen) K-edge absorption in
atoms, molecules, and solids, we refer the interested reader to a
recent review \citep{Frati2020}. 

\subsection{The oxygen K-edge of \CO}
\label{subsec:theory_co2}
\begin{figure}
  \resizebox{\hsize}{!}{\includegraphics{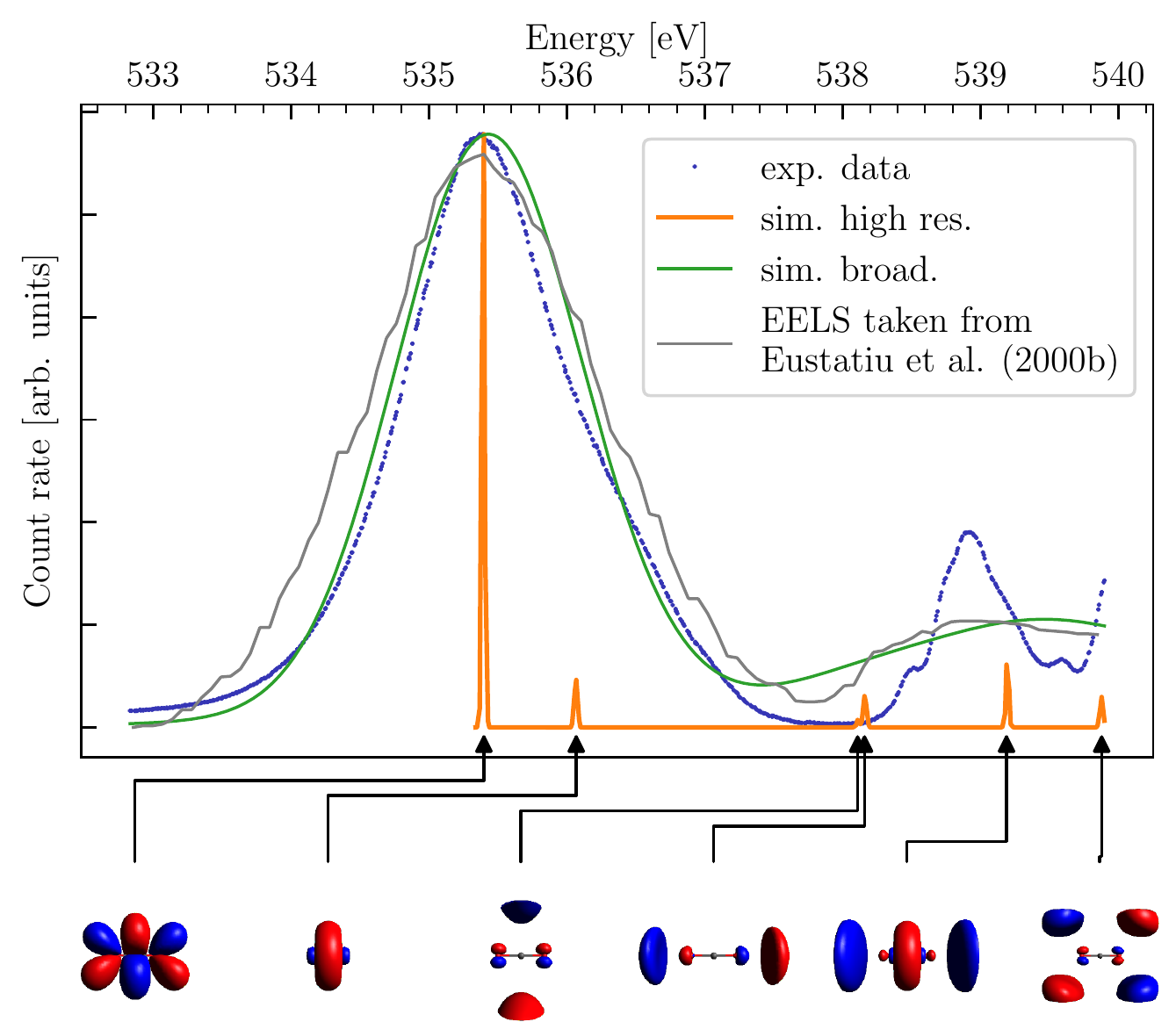}}
  \caption{\CO
    X-ray absorption spectrum from TDDFT calculations for oxygen
    K-edge. Highly resolved peaks were numerically broadened to visualize
    agreement of simulation and experiment. Experimental data are shown
    after subtraction of contamination. Additional data extracted from
    \cite{eustatiu2000b} with aligned first peak are depicted for
    comparison. The baseline of all of these spectra has been unified. Corresponding molecular orbitals are plotted below.}
    \label{fig:co2_simu}
\end{figure}

In Fig.~\ref{fig:co2_simu} we show the comparison of simulation
(orange and green lines) and experiment together with the earlier
 EELS data \citep{eustatiu2000b}. The first part of the
K-edge is dominated by the well known transition into the $\pi_u$
orbital at around 535.4\,eV. The peaks at higher energies are
typically assigned to Rydberg transitions. In this energy region we
get a satisfactory agreement in terms of the overall relative spectral
weight at a low resolution (see broadened simulation vs.\ EELS in
Fig.~\ref{fig:co2_simu}). However, the simulation misses the splitting
of peaks picked up by higher resolution experiments. An explanation
might be our neglect of vibrational modes. Indeed, \CO as a linear
($D_{\infty h}$) molecule, is a prime candidate even for irregular
vibrational structure due to the Renner-Teller effect
\citep{Frati2020}. Moreover, we point out that the simulated spectra
correspond to absorption with unpolarized light and can thus only be
directly compared to the EELS data \citep{eustatiu2000b}. We do not
account for matrix elements in the transition that account for
polarization dependence in the new experimental data. 

The summary of our \emph{ab initio} symmetry classification of the
peaks can be found in Table~\ref{tab:co2_theory}.

\begin{table}
    \caption{Assignment of irreps for transition orbitals with
      corresponding transition energy in K-edge excitation of \CO.}
    \label{tab:co2_theory}
    \begin{tabular}{lr}
        \hline \hline
        irrep  & energy [eV] \\
        \hline
        $e_{1u}$  ($\pi_u$) & 535.4\\
        $a_{1g}$  ($\sigma_g^+$) & 536.07\\
        $e_{1u}$  ($\pi_u$) & 538.11\\
        $a_{2u}$  ($\sigma_u^-$) & 538.16\\
        $a_{1g}$  ($\sigma_g^+$) & 539.19\\
        $e_{1g}$  ($\pi_g$) & 539.88\\
        \hline
    \end{tabular}
\end{table}

\subsection{The fluorine K-edge in \SF}
\label{subsec:theory_sf6}

\begin{figure*}
  \resizebox{\hsize}{!}{\includegraphics{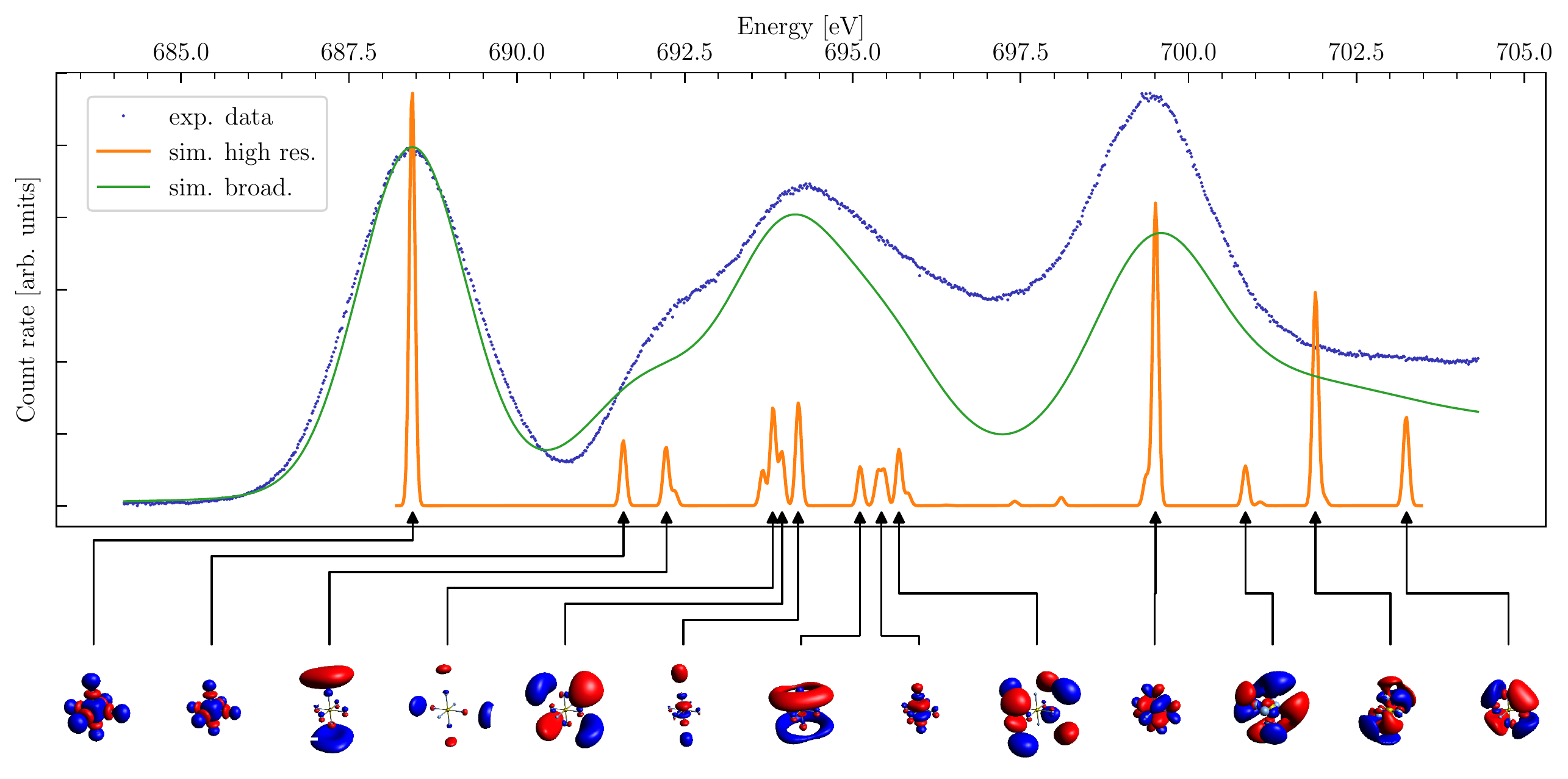}}
  \caption{SF\textsubscript{6}
    X-ray absorption spectrum from TDDFT calculations for fluorine
    K-edge. Highly resolved peaks are numerically broadened to visualize
    agreement of simulation and experiment. Experimental data are
    shifted to the baseline. The corresponding molecular orbitals are
    plotted below the $x$-axis.}
    \label{fig:sf6_simu}
\end{figure*}

In Fig.~\ref{fig:sf6_simu} we compare our simulation to the
experiment. In contrast to the oxygen K-edge of \CO, the fluorine
K-edge in octahedral ($O_h$) \SF is not dominated by a single
transition and has a comparable spectral weight in three main
structures between 685\,eV and 705\,eV. With the same calculation
parameters that we used for \CO, we find an overall satisfactory
agreement with experiment. The comparison reveals that particularly
the central double-peak structure around 693.5\,eV might originate
from a variety of transitions which are, however, not resolved in
experiment. In Table~\ref{tab:sf6} we provide a comprehensive list of
energies and symmetry character of the transitions.

\section{Conclusions}\label{sec:conclusions}

We used a newly introduced experimental setup to provide precise
calibration references in the soft X-ray regime. A careful statistical
analysis shows that the resulting energy calibration can in principle
provide an accuracy of 1--10\,meV (at energies in the 500--800\,eV
range). The resulting calibrations have no dependence on previous
measurements and therefore do not carry any legacy uncertainty present
in other measurements. The achieved accuracy is limited by significant
systematic uncertainties that exceed the statistical uncertainties by
almost an order of magnitude.
We performed several measurements of molecular absorption spectra that
are commonly used for energy calibration. The results for \CO show
relatively good agreement with previous ones; for \SF, we see a shift
similar to that found in \citep{leutenegger2020}. Significant
differences appearing in the \NE measurement compared to earlier works
might be partly an artefact of the non-simultaneous measurement of the
calibration, and require further investigation. Our theoretical
simulations of the \SF spectrum, although consisting of numerous
features, also show fairly good qualitative agreement with the data. We
are able to attribute a much richer structure underlying the measured
spectral weight at 691--697\,eV, supporting \citep{hudson1993} in contrast
to other works \citep{eustatiu2000a,francis1995,hitchcock1978}. For \CO,
the experimental spectra exhibit several features not captured by the
simulations.
We attribute these differences to us neglecting  polarization dependence
(dichroism) and vibrational effects. Since such vibrational and
symmetry-resolving effects do not influence the relative positions of
the peaks due to optical excitations, explicitly correlated methods
from many-body perturbation theory \citep[e.g.,
Bethe-Salpeter formalism;][]{Onida2002} may improve predictions from
theory.

Despite the systematic effects still present in our current
experiment, we have reduced the overall uncertainty in comparison with
various previous measurements. For further investigations aiming at
reaching a statistically dominated accuracy, it is necessary to follow
in time small relative energy shift of the photon beam energy selected
by the monochromator. This could be achieved by, e.g., photoemission
spectroscopy simultaneously performed with the photoionization
measurements, and would remove any dependency caused by the beamline.

The accuracy of theoretical calculations for few-electron HCI
surpasses that of any other soft X-ray standards, and thus our method
can in principle provide references at the level of $\sim$50\,meV
for this range. Such references will find various applications in
different fields of research, and help, as shown in this work,
assessing the accuracy of calculation for molecular systems.
Crucially, our calibration method and the present results address
essential needs of upcoming X-ray astrophysics missions.

\begin{acknowledgements}
Financial support was provided by the Max-Planck-Gesellschaft (MPG)
and Bundesministerium f\"ur Bildung und Forschung (BMBF) through
project 05K13SJ2. We thank HZB for the allocation of synchrotron
radiation beamtime at BESSY II. C.S.\ acknowledges the support by
an appointment to the NASA Postdoctoral Program at the NASA Goddard
Space Flight Center, administered by Universities Space Research
Association under contract with NASA, by the Lawrence Livermore
National Laboratory (LLNL) Visiting Scientist and Professional
Program Agreement No.\ VA007036 and VA007589, and by
MPG. Work by UNIST was supported by the National
Research Foundation of Korea (No.\ NRF-2016R1A5A1013277). Work by LLNL
was performed under the auspices of the U. S. Department of Energy
under Contract No.\ DE-AC52-07NA27344 and supported by NASA grants to
LLNL. M.A.L.\ and F.S.P.\ acknowledge support from NASA's Astrophysics
Program. Work by G.B.\ was supported by a NASA Space Technology
Research Fellowship. Work by R.C.\ was supported by NASA under award
number 80GSFC21M0002 and by an appointment to the NASA Postdoctoral
Program at the NASA Goddard Space Flight Center. Work by M.W.\ was
supported by the ERASMUS+ traineeship program.
\end{acknowledgements}

\section*{Author contributions}

J.S., M.W., N.H., P.H., J.W., and M.A.L.\ wrote the initial draft of the paper. The project was
initiated by M.A.L., J.R.C.L-U., S.B., and G.V.B. 
Theoretical calculations were carried out by M.W.\ and P.H. Data reduction and analysis was done
by J.S.\ and M.A.L. Everyone, except M.W.\ and P.H., was involved preparing the project
or conducting the experiment.

\end{document}